%% file: h0434_mn.tex
\documentstyle{l-aa}
\newcommand{\frat}[2]{\mbox{$\frac{
\raisebox{.1ex}{$\textstyle #1$}}{\raisebox{-.3ex}{$\textstyle #2$}}$}}

\newcommand{\fradd}[2]{\mbox{$\frac{\textstyle d #1}{\textstyle d #2}$}}
\newcommand{\MS}{\mbox{M$_{\textstyle \odot}$}}

\newcommand{\xmn}[2]{\mbox{$#1\!\times\! 10^{#2}\,$}}
\newcommand{\ergs}{\mbox{erg s$^{-1}$}}

\newcommand{\erggs}{\mbox{erg g$^{-1}$s$^{-1}$}}

\newcommand{\gccm}{\mbox{g cm$^{-3}$}}

\newcommand{\gsim}{\:\raisebox{.25ex}{$>$}\hspace*{-.75em}
      \raisebox{-.93ex}{$\sim$}\:}
\newcommand{\lsim}{\:\raisebox{.25ex}{$<$}\hspace*{-.75em}
      \raisebox{-.93ex}{$\sim$}\:}

\newcommand{\avr}[1]{\mbox{$<\!#1\!>$}}
\newcommand{\ewi}[1]{\mbox{Eq. \hspace*{-0.6ex}(\ref{#1})}}
\newcommand{\isn}[2]{\mbox{$^{#2}${#1}}}
\newcommand{\indis}[1]{{\mbox{\scriptsize #1}}}
\newcommand{\fref}[1]{\mbox{Fig.$\!$~\ref{#1}}}
\newcommand{\Avg}{\mbox{$N_\indis{A}$}}
\newcommand{\grpicture}[1]
{
        \epsfxsize=500pt
        \epsfysize=0pt
        \vspace{-90mm}
        \parbox{\epsfxsize}{\epsffile{#1.ps}}
        \vspace{-25mm}
}
\input epsf
\begin{document}
\thesaurus{02.14.1; 08.19.4}
\title{The neutrino-induced neutron source in helium shell and
r-process nucleosynthesis.}

\author{D.K.Nadyozhin \inst{1} \and I.V.Panov\inst{1,2}
\and S.I.Blinnikov\inst{1}\\}
\institute{Institute for Theoretical and Experimental Physics,
  B. Cheremushkinskaya St. 25, 117259, Moscow, Russia\\
\and
  Max-Planck-Institut f\"ur Astrophysik,
 Karl-Schwarzschild-Strasse 1, Postfach 1523,
  D-85740 Garching, Germany
  }

\date{}

\maketitle
\begin{abstract}

 The huge neutrino pulse that occurs
 during the collapse of a massive stellar core,
 is expected to contribute to the origination of a number of isotopes
 both of light chemical elements and heavy ones.
 In particular, evaporation of neutrons from helium nuclei excited
 by neutrino-nuclear inelastic collisions, may result in the r-process
 as it was first discussed by Epstein et al. (1988).
 Here we consider mainly the possibility to obtain
 the considerable amount of neutrons owing to the neutrino breakup of
 helium nuclei. It is shown that, in general,
 the heating of stellar matter due to
 the neutrino scattering off electrons and the heat
 released from the neutrino-helium breakup followed by the
 thermonuclear reactions should be taken into account.
 On the base of kinetic network, using all the important reactions
 up to $Z=8$, the main features and the time-dependent character
 of the neutrino-driven neutron flux are investigated.

  The time-dependent  densities of free neutrons produced in helium
  breakup,  $Y_\indis{n}(t)$,
  were used to calculate the r-process nucleosynthesis
  with another full kinetic network for $\sim 3200$ nuclides.
  It was found that in the case of
  metal-deficient stars, ${Z}\lsim 0.01\,{Z}_{\textstyle\odot}$, the
  resulting density of free neutrons seems to be high enough
  to drive the r-process efficiently under favorable conditions.
  But it is impossible to obtain a sufficient amount of
  heavy nuclei in neutrino-induced \mbox{r-process}
  in a helium shell at radii $R\! >\! R_\indis{cr}\approx 10^9$cm.
  We speculate that to make the neutrino-induced r-process work
  efficiently in the shell, one has to invoke nonstandard presupernova
  models in which helium hopefully is closer to the collapsed core
  owing, for instance, to a large scale mixing or/and rotation and
  magnetic fields.
  Apart from this exotic possibility, the neutrino-induced nucleosynthesis
  in the helium shell is certainly not strong enough to explain
  the observed solar r-process abundances.

 \end{abstract}

 \section{Introduction}
 About 20 years ago the gravitational
 collapse of massive stellar cores was recognized to be accompanied
 by a strong neutrino pulse, and a new branch of nucleosynthesis
 was born --- the neutrino nucleosynthesis. The general scheme of the
 neutrino nucleosynthesis can be described briefly as follows
 (e.g., Nadyozhin, 1991). Electron neutrinos $\nu_e$ and
 antineutrinos $ \tilde{\nu_e} $ can interact with different
 isotopes $(A,Z)$ of chemical elements by means of both charged
  currents:
\begin{equation}
  \nu_e + (A,Z)  \rightarrow (A,Z+1)^* + e^-  \, ,
  \label{NueHe}
\end{equation}
\begin{equation}
 \tilde{\nu_e} + (A,Z)  \rightarrow (A,Z-1)^* + e^+  \, ,
  \label{AnueHe}
\end{equation}
and neutral ones:
\begin{equation}
  \nu + (A,Z)  \rightarrow (A,Z)^* + {\nu}^{\prime}  \, ,
  \label{NMTHe}
\end{equation}
 where $\nu = \nu_e, \tilde{\nu_e}, \nu_{\mu}, \tilde{\nu_{\mu}},
 \nu_{\tau}$ and $\tilde{\nu_{\tau}}$. Since the typical muon and tau
 neutrino and antineutrino energies are sufficiently less than
 the rest mass of muon and tau-lepton, $\nu_{\mu}, \tilde{\nu_{\mu}},
 \nu_{\tau}$ and $\tilde{\nu_{\tau}}$ interact only by means of
 neutral current. The resulting nuclei are generally produced
 in highly excited states indicated by asterisks. The states decay
 emitting basically neutrons, protons, and $\alpha$-particles.
 The products of reactions (\ref{NueHe})--(\ref{NMTHe})
 interact both with themselves and
 background nuclei to give rise to a number of issues of importance
 for the origin of chemical elements. The existence of the
 Gamow--Teller and Isobaric-Analog resonances is crucial for the
 efficiency of the neutrino nucleosynthesis (see Fuller \& Meyer, 1995;
 Aufderheide et~al., 1994; Panov, 1994, and references therein).

 The beginning of this new issue of nucleosynthesis was marked by
 papers of Domogatsky \& Nadyozhin (1977, 1978, 1980),
 Domogatsky \& Imshennik (1982), and Domogatsky et~al. (1978a,b) devoted
 to the neutrino-induced production of p-nuclei and a number
 of light isotopes such as \isn{Li}{7}, \isn{Be}{9}, and \isn{B}{11}.

 Epstein et al. (1988) (EHC hereafter)
 put forward an elegant idea that, under
 favorable conditions, the inelastic scattering of the muon and tau
 neutrinos off helium nuclei in the helium shell would be a good source
 of neutrons necessary to drive the r-process. They showed that the
 favorable conditions were low metallicity
 $(\sim 0.01\,\mbox{Z}_{\textstyle\odot})$ and not too large radius of the
 helium shell ($\sim\xmn{7}{8}$cm). Woosley et al. (1990) undertook a
 comprehensive study of the neutrino nucleosynthesis and demonstrated,
 in particular, that it was hard to reconcile EHC's statement
 with their supernova models  (Woosley \& Weaver, 1995) ---
 the radius of the helium shell was large in their models and the burst of
 nucleosynthesis stimulated by the shock wave, crossing over the helium
 shell in a few seconds after the collapse of a stellar core, was
 strong enough (according to their estimate) to reduce the importance
 of the previous work done by neutrinos. The astrophysical community
 gives credit to Weaver \& Woosley for their presupernova models
 which are among the best nowadays. However, it would be premature
 to discard EHC's idea for the only reason that it does not fit
 these models.
 Although a major breakthrough in the understanding of the mechanism
 of Type II supernovae has been attained during last years,
 one cannot yet judge with full confidence such details of
 presupernova structure as the characteristics of chemical
 stratification just before and immediately after the onset of the
 neutrino pulse. The main uncertainties seem to be associated with
 the large scale properties of time-dependent convection
 (Bazan \& Arnett, 1994)  and with yet poorly studied effects of
  rotation at terminal stages of stellar evolution. It is plausible
  to imagine, for instance, the occurrence of aspherical
 hydrodynamic flows bringing helium closer to the stellar
 center than the radius of helium shell in spherically symmetric
 hydrostatic presupernova models.

Though a promising site for the
r-process nucleosynthesis responsible for the solar r-process abundances
is expected to be the neutrino-driven
wind blowing for a few seconds from a new-born neutron star
(Takahashi et al., 1994; Qian \& Woosley, 1996; Hoffman et al.,
1997), the possibility to create heavy elements in the helium shell
still remains to be of interest. First of all this relates to a well
developed r-process signature observed recently in the extremely metal
poor stars (Cowan et al., 1996a, 1996b; Ryan et~al., 1996;
Sneden et al., 1996) and possibly to some special issues requiring
the weak neutron fluxes such as the origin of some isotopic anomalies
in meteorites (Clayton 1989).

 The aim of our study
 in progress is to specify the conditions (in terms of helium
 shell radius, temperature, density, and metallicity) favorable
 for the neutrino-induced production of both light isotopes and
 free neutrons. Here preliminary results are reported on free
 neutron yields relevant to the r-process, which were already
 discussed briefly in Nadyozhin et al. (1996) and
 Nadyozhin \& Panov (1997).

\section{Properties of the neutrino flux and
                 thermonuclear rates}

\begin{figure}
\epsfysize=20cm
\epsfxsize=16cm
\vspace{-90mm}
\epsffile{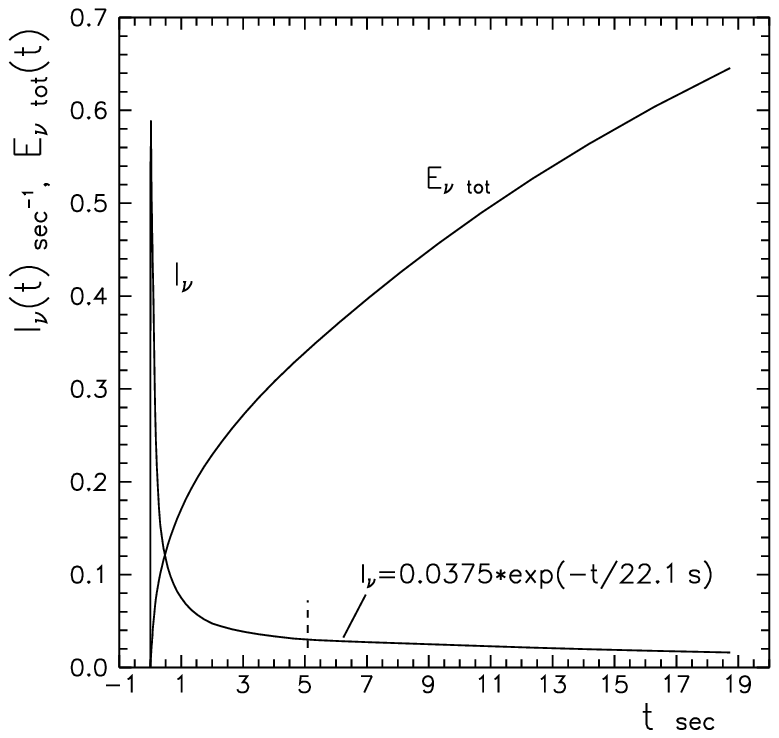}
\vspace{-55mm}
\caption{\noindent   The normalized neutrino light curve
  $l_\nu(t)=L_{\nu\tilde\nu}/{\cal E}_{\nu\tilde\nu}$}
 {and integrated energy of the neutrino flux
 $E_{\nu tot}(t)=\int_0^t\!l_\nu\, dt$, $E_{\nu tot}(\infty)=1$.
 (${\cal E}_{\nu\tilde\nu}$ is the total energy emitted
  by neutrinos of all the flavors).
       }
\label{lnbol}
\end{figure}

 We use the neutrino light curve calculated by Nadyozhin (1978)
 for the collapse of iron-oxygen star of 2\MS\ (\fref{lnbol}).
 The total energy emitted by neutrinos and antineutrinos of all
 the flavors is equal to ${\cal E}_{\nu\tilde\nu}=\xmn{5.3}{53}$erg.

 To obtain the neutrino luminosity in  \ergs and total energy emitted
 by neutrinos at time $t$ one has to multiply $l_\nu(t)$ and $E_\nu(t)$
 by ${\cal E}_{\nu\tilde\nu}$.
 The neutrino light curve has a narrow
 maximum at $t=0.023\,$sec and a slowly decaying ($\tau\sim 22\,$sec)
 long tail. The bulk of the available neutron star binding energy is radiated
 for as long time as $\sim 20\,$sec rather than for 3 sec adopted
 by EHC and Woosley et al. (1990).

 We assume that each neutrino flavor carries away
 the same energy equal to $\frac{1}{6}\, {\cal E}_{\nu\tilde\nu}$
 and has the Fermi--Dirac thermal spectrum with zero chemical potential.
 The mean individual energies of electron neutrinos and antineutrinos
 are taken to be equal to 12 MeV ($T_{\nu e}=3.81\,$MeV) whereas for muon and
 tau neutrinos and antineutrinos we have chosen a moderate value
  25 MeV ($T_{\nu\mu\tau}=7.94\,$MeV).
 The cross section of excitation of \isn{He}{4} by muon and tau neutrinos
 and the branching ratios for proton and neutron emission were
 calculated by EHC and Woosley et al. (1990) for the neutrino
 temperatures $T_{\nu\mu\tau}=4-12\,$MeV. We approximated
 their mean cross section (per helium nucleus and per neutrino
 flavor) by a simple equation
 \begin{equation}
 \avr{\sigma_{\indis{He4}\nu}}=\xmn{5.28}{-43}
  T_{\nu\mu\tau}^2\exp(-29.4/T_{\nu\mu\tau})\;\mbox{cm}^2\, ,
  \label{CROSS}
 \end{equation}
 where $T_{\nu\mu\tau}$ is in MeV. Equation (\ref{CROSS})
 reflects the threshold nature of the process and has
 an accuracy  better than 10\%
 within all the temperature range involved (4--12 MeV).

 The number of \isn{He}{4} nuclei destroyed by neutrinos per unit time
 is given by
 \begin{eqnarray}
  \dot{Y}_\indis{He4}  = - B(t)\, Y_\indis{He4}\, ;
 \quad B(t)\equiv\, q\,\frat{L_{\nu\tilde\nu}(t)
  \avr{\sigma_{\indis{He4}\nu}}}
   {4\pi R^2\avr{E_\nu}}=  \label{Trudy} \\
 \xmn{5.76}{-2}\frat{l_{\nu}(t)}{R^2_8}\,\mbox{s}^{-1},\; \quad
     \left(R_8=R/10^8\mbox{cm}\right)\, , \nonumber
 \end{eqnarray}
 where $q=2/3$ is the fraction of muon and tau neutrinos and
 antineutrinos in the total neutrino flux;
 $L_{\nu\tilde\nu}$ is the total neutrino-antineutrino luminosity;
 $R$ is the helium shell radius;
 and \avr{E_\nu} is the mean energy of
 muon and tau neutrinos and antineutrinos.
 The number of \isn{He}{4} nuclei in unit volume $n_\indis{He4}$ is
 given by $n_\indis{He4}=$$\rho \Avg Y_\indis{He4}$,
 \Avg being  Avogadro's number. So, $Y_\indis{He4}$
 is the number of \isn{He}{4} nuclei per baryon.
 The  emission rates
 of neutrons and protons (and respectively of \isn{He}{3} and
 \isn{H}{3}) are
 \begin{equation}
 \dot{Y}_\indis{n} =\dot{Y}_\indis{He3} =
     -b_\indis{n} \dot{Y}_\indis{He4}\, ,\qquad\qquad
 \dot{Y}_\indis{p} =\dot{Y}_\indis{H3}=
     -b_\indis{p} \dot{Y}_\indis{He4}\, ,\label{npdot}
 \end{equation}
 where $b_\indis{n}=0.471$ and $b_\indis{p}=0.516$ are the branching
 ratios (Woosley et al., 1990).

  To follow the sequence of thermonuclear reactions stimulated by
  the sources of fresh n, p, \isn{He}{3}, and \isn{H}{3}, we  have used
  a code connecting 26 nuclides from neutrons and protons up to \isn{O}{16}
  with a net of about 100 most important direct and inverse reactions
  presented in Tables 1 and 2.
  The reaction rates were taken mostly from Caughlan \& Fowler (1988)
  and from the updated version of Thielemann's original compilation
  (Thielemann et al., 1986).

\input h0434_in

  The equations of nuclear kinetics
  were calculated both for constant values of density and temperature and
  for time-dependent temperature induced by the heating of stellar matter
  owing to the neutrino-helium interaction and neutrino-electron scattering.

 \section{Neutrino heating of stellar matter}
  There are two sources of heat worth to be accounted for in the
  helium shell. They are the neutrino scattering off electrons and
  the heat
  released during the neutrino-helium breakup and the subsequent
  thermonuclear reactions.

 \subsection{Neutrino-electron scattering}

  The differential cross section of the neutrino scattering off
  electrons at rest ($E_\nu\gg kT$), is given
  by (e.g., Okun, 1982; Vogel \& Engel, 1989)
 \begin{eqnarray*}
 \fradd{\sigma}{\varepsilon} = S_0\left[\left(g_V+g_A\right)^2+
 \left(g_V-g_A\right)^2\left(1-\frat{\varepsilon}{E_\nu}\right)^2 +
 \right. \nonumber
 \end{eqnarray*}
 \begin{equation}
 \qquad{}  \qquad{}
 \left(g_A^2-g_V^2\right)
  \left. \frat{\varepsilon}
{E_\nu^2}
\right]\, ,\label{sigdf}
 \end{equation}
  where $S_0=\xmn{2.20}{-45}$cm$^2$, $E_\nu$ and $\varepsilon$ are
  the energy of incident neutrino and the recoil kinetic energy
  of electron, respectively (both in terms of $m_ec^2$), and
 \begin{eqnarray}
  g_V & = & \left\{\begin{array}{ll}
  2\sin^2\Theta_\indis{W}\, + \frac{1}{2}\qquad\mbox{ for }\nu_e ,\\
  2\sin^2\Theta_\indis{W}\, - \frac{1}{2}\qquad\mbox{ for }
  \nu_\mu,\,\nu_\tau ,\\
 \end{array}
 \right. \nonumber\\[-0.2cm]
  & & \label{gVgA}\\[-0.2cm]
  g_A & = & \left\{\begin{array}{ll}
 \phantom{-}\frac{1}{2}\qquad\qquad\qquad\quad\mbox{for }\nu_e ,\\
 -\frac{1}{2}\qquad\qquad\qquad\quad\mbox{for }\nu_\mu,\,\nu_\tau .\\
 \end{array}
 \right.\nonumber
 \end{eqnarray}
 Here $\Theta_\indis{W}$ is the Weinberg angle
 ($\sin^2\Theta_\indis{W}=0.23$).
 For antineutrinos, one must make the substitution
 $g_A\rightarrow -g_A$.
 The kinetic energy $\varepsilon$ depends on the angle
 of scattered neutrino and varies  within the limits
 \begin{equation}
  0\le\varepsilon\le\varepsilon_m=\frat{2E_\nu^2}{1+2E_\nu}
  \, .\label{elim}
 \end{equation}
 Integrating $\varepsilon d\sigma/d\varepsilon$  over
 the electron kinetic energy in the limits
 given by \ewi{elim} and averaging over the Fermi--Dirac
 neutrino spectrum, we obtain
 \begin{eqnarray*}
 \left<\!\varepsilon\fradd{\sigma}{\varepsilon}\!\right>\, =\,
 S_0 \sum_iq_i\left(\frac{F_4(0)}{F_3(0)}\frat{kT_i}{m_ec^2}\,
   a_i-b_i\right),
\end{eqnarray*}
 \begin{equation}
\quad {} \quad {} (i=\nu_e ,\tilde\nu_e ,\nu_\mu ,\tilde\nu_\mu ,
   \nu_\tau ,\tilde\nu_\tau )\, ,\label{sheat}
 \end{equation}
  where
  $$
   a_i=\frac{1}{12}\left(7g_V^2+7g_A^2+10g_Vg_A\right)_i,\quad
  $$
  $$
   b_i=\frac{1}{6}\left(5g_V^2+g_A^2+6g_Vg_A\right)_i,
  $$
  $q_i$ and $T_i$ are the fraction  in the total neutrino
  luminosity and the Fermi--Dirac temperature of type $i$ neutrinos,
  respectively; $F_4(0)$ and $F_3(0)$ are the Fermi--Dirac functions
  ($F_4(0)/F_3(0)=4.106$).
  Deriving \ewi{sheat}, we have neglected small terms of the order of
  $(m_ec^2/E_\nu)^2$.

 The rate of energy release per unit mass, $Q_s$ is given by
 \begin{equation}
 Q_s\, =\,\frac{1}{2}(1+\mbox{X})\Avg\frat{L_{\nu\tilde\nu}}{4\pi R^2}
 \left<\!\varepsilon\fradd{\sigma}{\varepsilon}\!\right>\, .
 \label{qheat}
 \end{equation}
 Here X is the mass fraction of hydrogen (in the helium shell X$=0$).

 Assuming $q_i=1/6$, we obtain numerically
  \begin{eqnarray*}
 Q_s\, =\,\xmn{1.267}{16}\frat{l_{\nu}(t)}{R_8^2} \cdot
  \end{eqnarray*}
  \begin{equation}
 \qquad{}  \left(0.4027\, T_{\nu e}+ 0.1730\, T_{\nu\mu\tau}-0.0657\right)
    \erggs\, ,\label{qsnum}
  \end{equation}
  where $T_{\nu e}$ and $T_{\nu\mu\tau}$ are in MeV and $R_8=R/10^8$cm.
  Since $T_{\nu\mu\tau}\approx 2T_{\nu e}$, the contribution of muon
  and tau neutrino and antineutrino scattering to heating is almost
  equal to that of the electron neutrino and antineutrino.

 \subsection{Neutrino-helium breakup and thermonuclear reactions}

  The neutrino-helium breakup supplies heat in the form of kinetic energy
  of the resulting fragments p, n, \isn{H}{3}, and \isn{He}{3},
  which
  is equal to the difference between the excitation energy $E^\ast$ and
  binding energy, $B_\indis{He4}$, of \isn{He}{4} with respect to
  $\isn{H}{3}+$p and $\isn{He}{3}+$n.
  Thus, the total release of heat
  per reaction is $E^\ast - B_\indis{He4}$ with
  $B_\indis{He4}\approx 20.6$ MeV and $\approx 19.8$ MeV for the
  proton and neutron channels, respectively. The general expression
  for the rate of energy release per unit mass due to
  the neutrino-helium breakup is given by

 \begin{eqnarray*}
  Q_B\, =\,\Avg\, Y_\indis{He4}\frat{L_{\nu\tilde\nu}}{4\pi R^2}\sum_iq_i
  \left[\avr{\sigma_{\indis{He4}i}}
  \frac{\left(E^\ast -B_\indis{He4}\right)_i}{\avr{E_{\nu i}}}\right],
 \end{eqnarray*}
 \begin{equation}
 \qquad{}  \qquad{} \,\,\, (i=\nu_e ,\tilde\nu_e ,\nu_\mu ,\tilde\nu_\mu ,
   \nu_\tau ,\tilde\nu_\tau )\, ,
  \label{QB}
 \end{equation}
 where $\left(E^\ast -B_\indis{He4}\right)_i$ is a branching-ratio
 averaged value and the Fermi--Dirac
 average energy of individual neutrinos $\avr{E_{\nu i}}= 3.147 kT_i$.

 To estimate the rate of energy liberation in the thermonuclear reactions,
 one has to find a difference between  the total binding energies
 of composition on two
 successive time steps of calculations, on line with
 the integration of the equations of nuclear kinetics.
 However, we shall use a simple approximation here. The point is that
 the thermonuclear reactions reassemble rapidly
 the bulk of the neutrino-destroyed helium and convert the rest of it
 into heavier nuclei (such as \isn{Li}{7}, \isn{B}{11}, \isn{C}{12}
 and others),
 thereby returning the binding energy of \isn{He}{4} back in the form of
 heat. To take  this effect into account, it is sufficient to neglect the
 binding energy $B_\indis{He4}$ in \ewi{QB}. Such a procedure
 slightly
 underestimates the energy release since the energy liberated
 in processing of \isn{He}{4} into heavier nuclei happens to be
 omitted.
 Assuming $B_\indis{He4}=0$ in \ewi{QB}, neglecting the electron
 neutrinos and antineutrinos, specifying $Y_\indis{He}=1/4$,
 $E^\ast = 36$MeV (Woosley et al., 1990), and $\sum q_i=2/3$, and using
 \ewi{CROSS} for \avr{\sigma_{\indis{He4}i}}, we obtain numerically

 \begin{eqnarray*}
  Q_B\, =\,\xmn{2.557}{18}\frat{l_{\nu}(t)}{R_8^2}\,
  T_{\nu\mu\tau}\exp(-29.4/T_{\nu\mu\tau})\, =\,
 \end{eqnarray*}
 \begin{equation}
\qquad {}  \xmn{5.017}{17}\frat{l_{\nu}(t)}{R_8^2} \; \erggs\, ,
  \label{QBN}
 \end{equation}
  where we made also a substitution $T_{\nu\mu\tau}=7.94$ MeV.

 \subsection{Neutrino heating}
 For simplicity we assume that density $\rho$ and radius $R$
 of the helium shell remain
 unchanged during the heating. This is certainly true for the
 initial narrow peak of the neutrino luminosity (\fref{lnbol}). However,
 the time scale of the neutrino light curve tail is comparable to the
 hydrodynamic time scale in the helium shell, $\sim R/v_{\indis{sound}}$,
 and such an assumption slightly overestimates the resulting increase
 in temperature.
 To obtain the temporal behavior of temperature,
 one has to integrate over time
 the thermodynamic equation of energy:
  \begin{equation}
  E(T,\rho)\, =\, E(T_0,\rho)\, +\, \int_0^t\! (Q_S+Q_B)\, dt\, ,
  \label{temp}
  \end{equation}
  where $E$ is specific energy.  For $E$, we use  the Equation of State
  as given by Blinnikov et~al. (1996) which takes into account ideal
  electron-positron gas with an arbitrary degree of degeneration,
  ideal ion gas, and blackbody radiation.
\begin{figure}
\epsfysize=20cm
\epsfxsize=16cm
\vspace{-65mm}
\epsffile{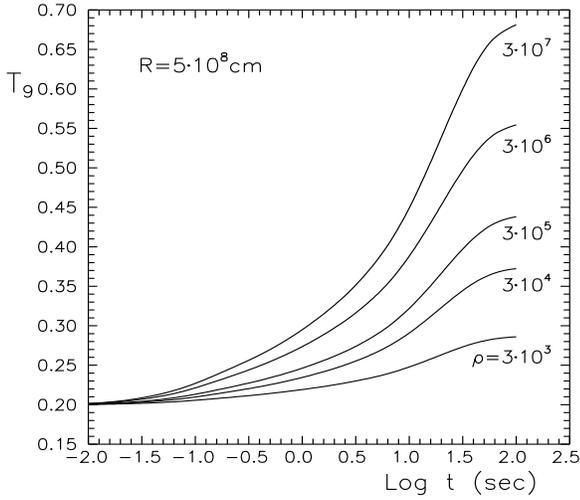}
\vspace{-70mm}
\caption{\noindent Temperature versus time as a result of
 the neutrino heating. The initial temperature $T_9(0)=0.2$.}
\label{heatnl}
\end{figure}

  Figure \ref{heatnl} shows temperature
  as a function of time for different $\rho$ as calculated
  with \ewi{temp}.
  For initial temperature
  $T_9=0.2$, density $\rho =\;$\xmn{3}{3}\gccm, and radius $R=\;$\xmn{5}{8}cm
  of the helium shell, the heating is noticeable but not crucial ---
  $T_9$ increases up to $\approx 0.25$ in 10 sec after the onset of
  the neutrino flux. The greater density, the stronger heating ---
  for instance, at $\rho =$\xmn{3}{5}\gccm, temperature increases
  up to 0.32. Since the heating scales also as $R^{-2}$, it must be
  taken into account in the calculations of the neutrino
  nucleosynthesis at radii $R\lsim$\xmn{5}{8}cm (especially for
  the carbon and silicon shells).

\begin{figure}
\vspace{+15mm}
\epsfysize=20cm
\epsfxsize=17cm
\vspace{-110mm}
\epsffile{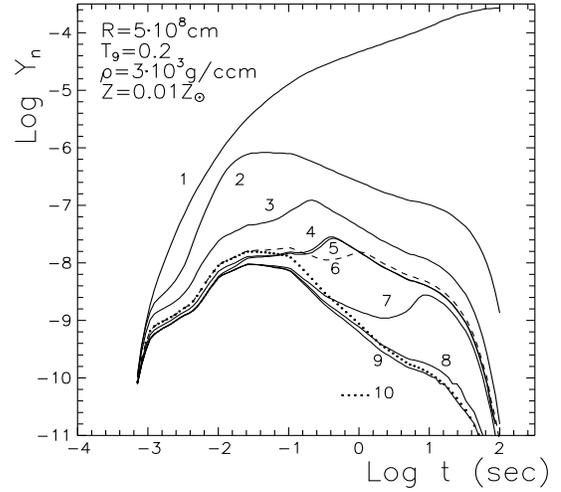}
\vspace{-40mm}
\caption{\noindent The neutron abundance versus time.}
{ The curves labelled by numbers correspond to the cases:

 \begin{tabular}{rl}
  1   & Neutron yield from the neutrino-helium inelastic \\
      & scattering (all the nuclear reactions being \\
      & turned off). \\
  2   & No "Fe" and no initial \isn{N}{14}. \\
  3   & No \isn{N}{14}, the burning rate of Fe is slowed by \\
      & a factor of 30.\\
  4   & No Fe and the CNO fraction of initial \isn{N}{14}. \\
  5   & Cosmic initial fraction of Fe and CNO fraction \\
      & of \isn{N}{14}; Fe burns via $^{56}$Fe(n,$\gamma$) reaction. \\
  6   & A half of CNO fraction of \isn{N}{14}, the burning rate \\
      & of Fe is slowed by a factor of 30 (dashed curve). \\
  7   & Same as Case 6 but CNO fraction of \isn{N}{14}.\\
  8   & The CNO fraction of \isn{N}{14} and burning rate of \\
      & Fe is slowed by a factor of 100. \\
  9   & Same as 5 but no burning of Fe (Y$_\indis{Fe}$=const).\\
 10   & Same as 6, but no burning of Fe (dotted curve).
 \end{tabular}

}
\label{nnoht1}
\end{figure}

 \section{Neutrino-induced production of neutrons}

 A number of calculations were made for different (including extreme)
 assumptions of initial amounts of iron and \isn{N}{14} and of
 the helium shell radius $R$, as well.
 The temporal behavior of neutron yield $Y_\indis{n}$
 is shown in \fref{nnoht1} for $R=\xmn{5}{8}$cm, $T_9=0.2$,
 $\rho=\xmn{3}{3}$ g cm$^{-3}$, $\mbox{Z}=0.01\mbox{Z}_{\textstyle\odot}$,
 and different assumptions about the neutron poisons \isn{N}{14}
 and \isn{Fe}{56}.
 The neutron density  $n_\indis{n}(t)=$
 $\rho \Avg Y_\indis{n}(t)=\;$$\xmn{1.81}{27}Y_\indis{n}(t)\,
 \mbox{cm}^{-3}$.

  Even with no initial \isn{N}{14} and Fe (Case 2), the neutron density
  falls substantially (by about 4 orders of magnitude during the first 10 s)
  as compared to the amount supplied by the
  neutrino-helium inelastic scattering (Case 1)
  \footnote{According to Eqs. $\!$(\ref{Trudy},\ref{npdot}), the total
   amount of
   neutrons produced by the neutrino breakup of \isn{He}{4} is equal to
   $Y_{\indis{n}}=\xmn{5.76}{-2}b_{\indis{n}}Y_{\indis{He4}}/R^2_8
   \approx\xmn{2.7}{-4}$ for $R_8=5$.}.
  This happens because
  the bulk of the neutrino-destroyed \isn{He}{4} proves to be reassembled
  mostly through  \isn{H}{1}(n,$\gamma$)\isn{H}{2},

\noindent  \isn{He}{3}(n,p)\isn{H}{3}, \
  \isn{H}{2}(D,p)\isn{H}{3}, \  \isn{He}{3}(D,p)\isn{He}{4},
  \  \isn{H}{3}(D,n)\isn{He}{4},

\noindent
 \isn{He}{3}(T,D)\isn{He}{4}, and \isn{H}{3}(T,2n)\isn{He}{4} reactions.
  Only as little as $\sim$1\%  of the \isn{He}{4} breakup products is
  left unbound back into \isn{He}{4}
  to synthesize some \isn{Li}{7} [mostly through the reaction
  \isn{He}{4}(T,$\gamma$)\isn{Li}{7}] and other less abundant
  species such as \isn{B}{11}, \isn{C}{13}, and \isn{C}{14}.

  In Case 4, there is no Fe and the
  initial fraction of \isn{N}{14} is equal to the cosmic fraction of
  CNO-isotopes multiplied by a factor of 0.95 as it follows from
  the hydrogen burning in well developed CNO-cycle
  $(Y_\indis{N14}=0.95\,\mbox{X}_\indis{CNO}/14= 0.0523 \mbox{Z})$.
  The same initial
  value of $Y_\indis{N14}$ was chosen also for Cases~5, 7, 8, and 9.
  For the versions described by Cases~5--9,
  the initial abundance of Fe was assumed to be equal to its cosmic
  value of $Y_\indis{Fe56}= \mbox{X}_\indis{Fe}/56=\xmn{1.27}{-3}\mbox{Z}$
  (Anders \& Grevese, 1989). The difference between Cases~5 and~4
  is that in Case 5 the initial abundance of Fe was taken equal to its cosmic
  value but Fe was burning at the rate
  \begin{equation}
  \fradd{Y_\indis{Fe}}{t}\, =\,
  - \rho \Avg\avr{\sigma v}_{\indis{Fen}\gamma}
   Y_\indis{n}Y_\indis{Fe}\, .\label{dFedt}
  \end{equation}
  Since the number of the neutrino-produced neutrons (Case~1) exceeds
  considerably the initial number of Fe-nuclei, it takes only about
  0.05 sec to burn out virtually all Fe. That is why one can hardly notice
  the difference between Cases~4 and~5.  This is, of course,
  an unreal assumption. The products of Fe burning still continue to
  absorb neutrons. To simulate this effect we calculated two more
  versions with the rate of Fe-burning decreased by factors 30 and
  100 (Cases~7 and~8, respectively).
  In other words, we  have changed \ewi{dFedt} by multiplying its right-hand
  side by an additional factor of 1/30 or 1/100. However, in the equation
  for $dY_{\indis{n}}/dt$, the rate of the neutron absorption by Fe
  remained unchanged and equal to the unmodified right-hand side of
  \ewi{dFedt}.
  This means that $Y_\indis{Fe}$
  begin to decrease substantially after capturing 30 or 100 neutrons.
  The factor 1/30 in the right-hand side of \ewi{dFedt} seems to be
  a good compromise to simulate the consumption of neutrons in the
  r-process. In Case~9, Fe does not burn at all.
   In Cases 4--6, the neutron abundance reaches the magnitude
   of $Y_\indis{n}\sim 10^{-8}$ and keeps it as long as about 5 sec.
  The duration of the plateau in Cases 7--8
  is much shorter than for Cases 4--6. The comparison of Case~10 with
  Case~6 shows how strongly $Y_\indis{n}$ depends on the parameters
  of Fe-burning at $t\gsim 1\,$sec.

    The initial abundance of \isn{N}{14}, equal to the total amount
    of CNO isotopes, is certainly the upper limit.
    In moderately massive stars
    (10--15\MS ), the CNO-cycle is hardly to be efficient enough
    to convert all \isn{O}{16} into \isn{N}{14}. Thus, it is
    reasonable to take a lower initial abundance of \isn{N}{14}.
    In Case~6, $Y_\indis{n}$ was calculated with
    $Y_\indis{N14 initial}= 0.5\mbox{X}_\indis{CNO cosmic}/14$
    (which gives  $Y_\indis{N14}/Y_\indis{Fe}\approx 20$ --- the value
     accepted by EHC)
    and the rate of Fe-burning decreased by a factor of 30. This
    set for properties of the neutron poisons \isn{N}{14} and Fe
    will be considered as standard in our further preliminary r-process
    calculations.

\begin{figure}
\epsfysize=20cm
\epsfxsize=16cm
\vspace{-120mm}
\epsffile{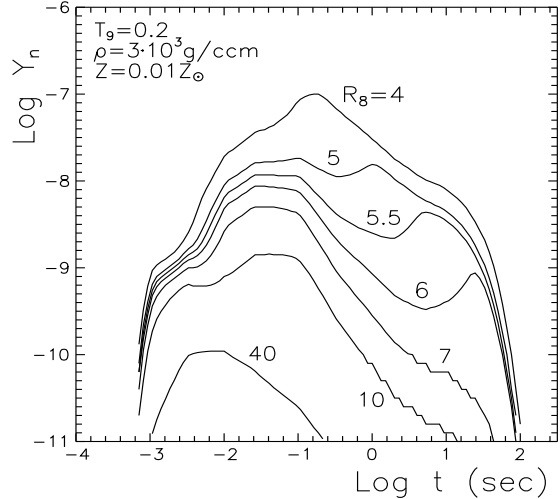}
\vspace{-10mm}
\caption{\noindent Neutron abundance versus time for different
          radii of helium shell.}
\label{nnohtr}
\end{figure}

    The resulting $Y_\indis{n}$ proved to be strongly sensitive to
    the helium shell radius $R$ (\fref{nnohtr}).
    For instance for $R=10^9$cm, $Y_\indis{n}$ attains a maximum
    $Y_\indis{n max}\approx 10^{-9}$ which
    is about an order of magnitude less than for
    $R=\xmn{5}{8}$cm and lasts only for $\sim 0.2\;$sec
    rather than $\sim 5\;$sec as in Case~6.

\begin{figure}
\epsfysize=20cm
\epsfxsize=16cm
\vspace{-125mm}
\epsffile{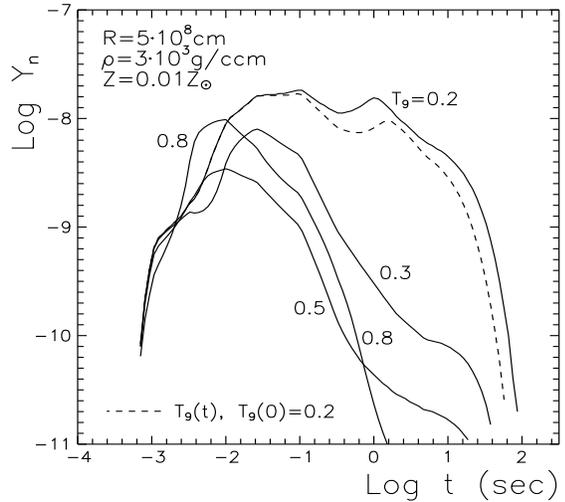}
\vspace{-5mm}
\caption{\noindent Neutron abundance versus time for different
          temperatures of helium shell.}
\label{nhtro33}
\end{figure}

  The dependence of $Y_\indis{n}$ on temperature turns out to be
  more complicated (\fref{nhtro33}).
  The solid lines correspond to constant temperatures
  $T_9=0.2$ (Case~6), $0.3$, $0.5$, and $0.8$ whereas the dashed one shows
  the effect of neutrino heating for initial temperature $T_9=0.2$ with
  $T_9(t)$ given by the lower curve in \fref{heatnl}.
  The neutron yield falls abruptly even for a moderate increase
  in $T$. However at temperatures $T_9\gsim 0.8$, it starts
  to grow at the very beginning of the neutrino flux
  $(t\lsim 0.1\;\mbox{sec})$.
  This happens owing to a considerable rise in the rates of (p,n),
  ($\alpha$,n), and ($\gamma$,n) reactions. The possibility of the
  neutrino-driven r-process at temperatures as high as $T_9\sim 2$
  seems to deserve further and more detailed investigation.

 \section{The r-process}
  The time-dependent neutron yield $Y_\indis{n}(t)$
  discussed above can now be used to calculate the
  r-process. Instead of rough approximations for
  $Y_\indis{n}(t)$ imitating Case~6
  in our previous calculations
  (Nadyozhin et~al., 1996, Versions I and II),
  we used the time-dependent neutron yield exactly as shown
  for  Case~6 in \fref{nnoht1} and its
  analog for $R=10^9\;$cm (the curve $R_8=10$ in \fref{nnohtr})
  to fulfill the r-process calculations.
  Such an approximate approach
  to connect the two codes, through the neutron abundance
  $Y_\indis{n}(t)$ seems to be adequate for our preliminary study.
  The accuracy of the approximation is briefly discussed in Appendix.

  The r-elements yields were estimated with the aid of the
  kinetic model of nucleosynthesis developed earlier by
  Blinnikov \& Panov (1996).
 In this model, variations in the
 number density $n(A,Z)$ of each nuclide were determined taking
 into account reactions involving neutrons and $\beta$-decays :
\begin{eqnarray}
& &   dY_{A,Z}/dt =
   -\lambda_{\gamma\mbox{\scriptsize n}}(A,Z) \cdot Y_{A,Z}
   - \lambda_{\beta} (A,Z) \cdot Y_{A,Z}
   - \nonumber \\
& & \phantom{dYA}Y_\indis{n}(t)\rho\Avg
   <\!\sigma_{\mbox{\scriptsize n}\gamma}(A,Z) v\!> Y_{A,Z} +
   \nonumber \\
& &\phantom{dYA}\lambda_{\gamma\mbox{\scriptsize n}}(A+1,Z)\cdot Y_{A+1,Z}+
     \nonumber \\
& &\phantom{dYA}  Y_\indis{n}(t)\rho\Avg
  <\!\sigma_{\mbox{\scriptsize n}\gamma}(A-1,Z) v\!>
   Y_{A - 1, Z} + \nonumber \\
& &\phantom{dYA}   \lambda_{\nu e} (A,Z-1)\cdot Y_{A,Z-1}
   - \lambda_{\nu e} (A,Z) \cdot Y_{A,Z}  +  \nonumber \\
& &\phantom{dYA}\sum_{i=0,1,2,3}\!\lambda_{\beta}(A+i,Z-1)\cdot P_i(A+i,Z-1)
      \label{abuaz}   \nonumber \\
& &\phantom{dYA} \cdot Y_{A+i,Z-1} \; ,\\
& &  dY_\indis{n}/dt = \sum_{A,Z}\!\!
 \left[\lambda_{\gamma\mbox{\scriptsize n}}(A,Z)-Y_\indis{n}(t)\rho
 \Avg\avr{\sigma_{\mbox{\scriptsize n}\gamma}(A,Z)v}
 \phantom{\sum_{A,Z}}\right.\nonumber\\
& & \left. +\sum_{i=1,2,3}\! i\cdot\lambda_{\beta}(A,Z)
   \cdot P_i(A,Z)\right] Y_{A,Z}\; , \label{Yndt}
 \end{eqnarray}
 where
 \begin{equation}
 Y_{A,Z} =\frac{n(A,Z)}{\rho\Avg} \; , \; \qquad
    Y_\indis{n}(t) = \frac{n_\indis{n}(t)}{\rho\Avg}\; .
 \label{igrek}
 \end{equation}
 In addition to equations given by Blinnikov \& Panov (1996),
 Eqs.~(\ref{abuaz},\ref{Yndt}) contain terms describing
 the beta-delayed emission of
 neutrons,  $P_i(A,Z)$ being the probability of
 emission of  $i$ neutrons, ($\sum_{i=0,1,2,3} P_i=1$).
  However, in the present calculations we take into account the beta-delayed
  emission of only one neutron, namely
  we assume $P_2=P_3=0$.
  Equation (\ref{Yndt}) for $dY_\indis{n}/dt$
 is not used here since $Y_{\indis{n}}(t)$ was taken
 from  Case~6 in \fref{nnoht1}.

 Moreover, we have also added terms  accounting for
 the capture of {\em electron\/} neutrinos by nuclei with
 reaction rates
 $\lambda_{\nu e}(A,Z)$.
 The relative role of the $\nu_e$-capture
  in \ewi{abuaz} turns out to be insignificant for the environments
  under consideration.
 The contribution of the
 $\tilde\nu_e$-capture is even smaller owing to a high threshold
 for the neutron-rich nuclei, and we neglect it.
 The neutrino-nuclear cross sections
 were calculated earlier (Panov, 1994; cf. Fuller \& Meyer, 1995;
 McLaughlin \& Fuller, 1995), with both
 Isobar-Analog and Gamow--Teller resonances taken into account.
The hope to obtain a noticeable smoothing of the abundance curve of the
r-process elements due to the neutrino-nuclear interaction, seems
hardly to be realized since the cross sections $\sigma_{\nu e}(A,Z)$
themselves demonstrate a clear even-odd effect that is decreasing,
however, in magnitude when $A$ and $Z$ are increasing.
The  r-process calculations, fulfilled here with account for
the capture of the {\em electron \/} neutrinos by nuclei,
manifested only one noticeable effect --- the acceleration
of the r-process, as it was shown by Nadyozhin \& Panov (1993).
However, in case of a larger contribution of the neutrino-capture
reactions to the r-process or under the conditions of lower neutron fluxes
involved (owing, for instance, to the exhaustion of neutrons at the
very end of the r-process), neutrinos would be able to decrease the
even-odd effect to some extent (Panov, in preparation).
It should be mentioned that our calculations do not take into account
the evaporation of nucleons from highly excited states of daughter nuclei.
This effect seems to be only of minor importance for the bulk of stable
nuclei synthesized in the r-process (Meyer et al., 1992) except
for the production of p-nuclei (Domogatsky \& Nadyozhin, 1977, 1978).

  It is quite appropriate to notice here that the calculations of
  $\sigma_{\nu}$ for neutrinos from collapsed stellar cores in the
  approximation of the GT-resonance energy shape with the delta
function (Fuller \& Meyer, 1995; McLaughlin \& Fuller, 1995), are
accurate enough only when $<\!E_{\nu}\!>$ $\approx$ $E_\indis{GTR}$.
However, in case of typically short-living nuclei along the r-process
path, the GT-resonance energy proves to be rather high (Panov, 1994)
and one has to deal with a detailed shape of GT-strength function in
more accurate calculations.

 The boundary conditions for the range of the nuclei involved into
 consideration, were specified in the following way:
 $Z_\indis{min}$=26 and $Z_\indis{max}$=95,
 with the minimum and maximum $A$ for each $Z$ being
 determined by the lightest stable nucleus from one side, and by the
 neutron-drip line for the heaviest isotope from the other side.
 Although we did not take into account the proton-rich unstable nuclei,
 the full number of nuclides and kinetic equations involved in our
 calculations was as large as
 about 3200 (the strict number depended on mass formulae used).
 The reaction rates entering Eqs.~(\ref{abuaz})
 differ by tens orders of magnitude.
 Thus, the system of equations for nuclear kinetics,
 to be dealt with, is a classical example of a stiff system of
 ordinary differential equations.
In our calculations, we used  one
of the most effective methods to solve such a stiff system of equations ---
Gear's method (Gear, 1971). Our code has options permitting
also to make use of the implicit Adams method and the
method by Brayton et~al. (1972).
The description of the complete package of solver routines
and its applications to the r-process calculations,
can be found in Blinnikov \& Panov (1996). A similar full network
  is developed by Goriely \& Arnould (1996).

 Nuclear reaction rates were taken mainly from Thielemann et~al. (1986),
 and mass relations from Hilf et~al. (1976), though we also considered
 the dependence of r-process yields on various
 mass relations and nuclear reaction sets.  Beta-decay rates
 were taken from Aleksankin et~al. (1981). We evaluated the influence of
 different sets of  beta-decay rates, either.
 The comparison with
 beta-decay rates of Staudt et~al. (1990) shows,
 that for the majority of nuclei with even $Z$, the calculated
 values are rather close to each other. Nevertheless,
 for some isotopic chains with
 odd $Z$, the deviations may be rather large. However, all these deviations
 may affect significantly only the isotopic yields but not the elemental ones.
 Here we concentrate
 mainly on the efficiency of the neutrino-induced r-process under the
 astrophysical conditions expected for the helium shell
 while our discussion of influence of
 different nuclear data available is not so detailed; in particular,
 it concerns the beta-decay rates, beta-delayed neutron emission
 probabilities, and the rates of the electron neutrino nuclear capture.
\begin{figure}
 \vspace{-105mm}
\grpicture{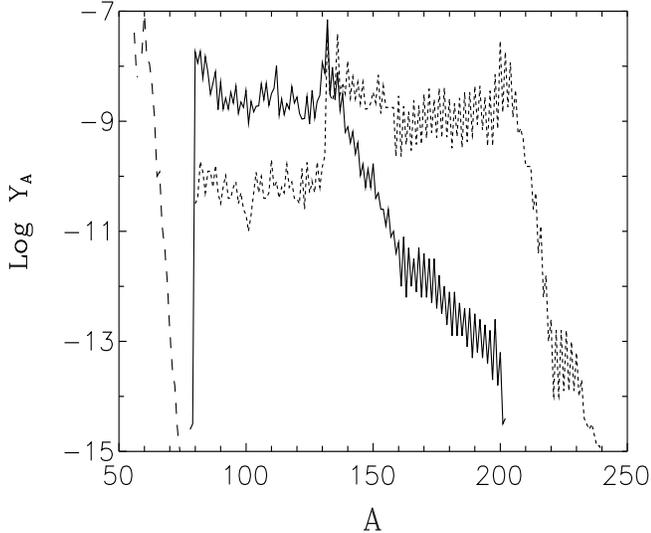}
\vspace{50mm}
\caption{\noindent The r-process abundances
   $Y_\indis{A}$ = $ \sum_Z Y(A,Z)$ as a function of atomic mass.
    Version I :
  $R=\xmn{5}{8}$cm,
  duration time $\tau=10\; $sec (solid line) and 35 sec (dotted line).
  Version II (dashed line):  $R=10^9\;$cm,
  duration time $\tau=100\; $sec.
   Mass formulae are from Hilf et~al. (1976).}
\label{fig3a}
\end{figure}
 Results of the calculations are shown in Figs. \ref{fig3a} and \ref{fig3d}
  for Versions I and II and for various mass formulae.
 For Version I,
 independent of the mass formulae used, the nuclei
 from $A=80$ through $A=130$ and then up to $A=200$
 were synthesized  by the $10$-th
 and $35$-th second, respectively.
 Under the  conditions of Version II it is possible also
 to form some r-process nuclei but not far from the iron peak.

 The differences in the shape of the abundance curves based on different data
 sets are not very high, except for the region of low $A$, where the initial
 nuclei burn out a little bit slower when the data of Hilf et~al.
 (1976) and reactions rates of Thielemann (1986) are used
 instead of combining the data of J\"anecke \& Eynon (1976)
 with reactions rates of Panov (1995, 1997)
(see Figs. \ref{fig3a} and \ref{fig3d}).

 In our network calculations, we have taken into account
 the emission of beta-delayed neutrons to study whether they are
 able to reduce the amplitude of the odd-even effect.
 The smoothing of abundances due to the emission of beta-delayed
 neutrons proved, however, to be small in our case, since
 the neutron flux is not very high and r-process
 nucleosynthesis occurs not far from the stable nuclei
 having small values of the beta-delayed neutron emission probabilities.
 This effect depends mostly upon the changes in r-process path
 after the break-down of nuclear statistical equilibrium and usage
 of the kinetic network of nucleosynthesis (Howard et~al., 1993).

 As it can be seen from Figs. \ref{fig3a} and \ref{fig3d},
 in the region of $A\gsim 150$
 the odd-even effect  is a little bit more pronounced -- maybe because of
 the boundary effects such, for example, as a step-like shape of
 the stable nuclei boundary (Panov \& Blinnikov, in preparation).
\begin{figure}
 \vspace{-37.mm}
\grpicture{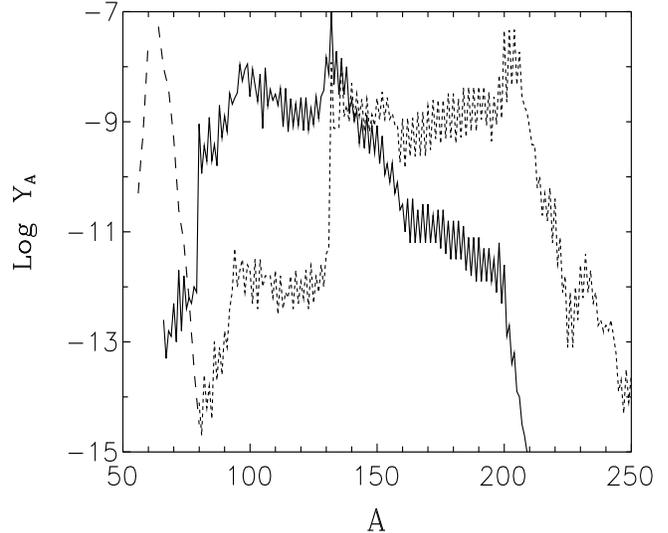}
        \vspace{-18.mm}
\caption{Same as in \protect\fref{fig3a}, but Mass formulae are taken from
  J\"anecke \& Eynon (1976), and nuclear reaction rates from
 Panov (1995, 1997).}
\label{fig3d}
\end{figure}

 An additional smoothing of the odd-even effect could occur
 due to emission of the beta-delayed neutrons during subsequent
 beta-decays into stable isotopes.
 But under the astrophysical conditions discussed here, the r-process path lies
 not far from beta-stable nuclides and only a fraction of the nuclei formed in
 sufficient quantity have significant values of beta-delayed neutron
 emission probabilities. That is why in the present case the smoothing of
 the odd-even effect due to delayed neutrons happens to be small
 and cannot change significantly the calculated abundance curves.

\section{Discussion and Conclusions}
  In this work we began a systematic study of the conditions favorable
for the neutrino-induced production of free neutrons in the amounts sufficient
to form heavy elements through the \mbox{r-process} in stellar shells
 surrounding the collapsed core --- a source of the intense neutrino flux.
 The problem in question is described by four main parameters: radius of
 the shell $R$ (the neutrino flux scales as $R^{-2}$), density $\rho$,
 and temperature $T$ in the shell, and chemical composition of the shell.
 The latter parameter splits in two important ones --- the main constituent
 (helium, carbon, silicon etc.) and the admixture of the neutron poison
 species like \isn{N}{14} and \isn{Fe}{56}, the latter being both a raw
 material for the r-process and a neutron poison as well.
 We concentrate here on the possibility of the r-process in a helium
 shell consuming neutrons from the breakup of \isn{He}{4} by neutrinos.

 We explored the dependence of the neutron yield, $Y_\indis{n}$, on $R$
 varying within \xmn{4}{8} -- \xmn{4}{9}cm (\fref{nnohtr}) for constant
 values of temperature, density, and metallicity:
 $T_9=0.2$, $\rho =\xmn{3}{3}$\gccm, and Z$=0.01\mbox{Z}_{\textstyle\odot}$.
 The calculations of the r-process with  $Y_\indis{n}(t)$ given by the
 curves for $R=\xmn{5}{8}$ and $10^9$cm in \fref{nnohtr} show that
 there exists a critical radius of the helium shell
 $R_\indis{cr}\approx 10^9$cm, above which the neutrino flux fails to
 stimulate the r-process efficiently.

 The calculations of $Y_\indis{n}$ for different temperatures of the
 helium shell (\fref{nhtro33}) demonstrated a strong sensitivity
 of $Y_\indis{n}$ to temperature. It was also shown that in
 accurate calculations one has to take into account the heating of
 helium shell by neutrinos, especially if the density is large
 enough $\left(\rho\gsim 10^5\gccm\right)$. For the initial temperature
 $T_9\gsim 0.5$, $Y_\indis{n}$ begins to increase with temperature,
 first at the very onset of the neutrino flux and then progressively
 at longer times (compare curves $T_9=0.5$ and $0.8$ in \fref{nhtro33}).
 If the initial temperature of the helium shell happens to be as large as
 $T_9\approx 2$, one can expect a certain increase in the critical radius
 $R_\indis{cr}$. From one side, such temperatures prevent
 the neutrino-destroyed helium to be reassembled and thereby support
 $Y_\indis{n}$ at a higher level. From the other side,
 at such a high temperature,  thermonuclear
 reactions ($\gamma$,p), ($\gamma$,n), (p,n), ($\alpha$,n) become
 fast enough  to contribute themselves to  $Y_\indis{n}$ even when
 the neutrino flux is absent (Woosley et al., 1990).

 It is worth noticing
 that the shock wave, passing across the helium shell in a few seconds
 after the collapse of stellar core, heats matter up to $T_9=2$--$3$
 and reprocesses the products of the neutrino-induced nucleosynthesis.
 This effect deserves more detailed study. Here we would like only
 to mention an interesting chance (see also Domogatsky et al., 1978b)
  given by a possibility of collapses not leading to standard
 powerful supernova explosions. This chance can be high if, e.g., the
 neutron star formation rate is higher than the rate of collapsing
 supernova outbursts. Unfortunately, the uncertainties of observations
 do not permit to prove this to be true,
 yet they do not reject this chance, either.
 It is conceivable, that  the rotation of a presupernova plays
 the major role during those almost silent collapses.
 So, a large mass of the material
 would be kept at the radius lower than $R_\indis{cr}$ by the angular
 momentum conservation. Later, as suggested by Bisnovatyi-Kogan (1970, 1980),
 the magnetic field, enhanced by the differential rotation, can eject the
 outer layers (already irradiated by the neutrino flux). It is interesting
 to note that the latest computations (Bisnovatyi-Kogan et~al., 1995)
 fail to produce an explosion on the supernova energy scale. So,
 a powerful shock wave, able to dissociate the r-process nucleosynthesis
 yields, is not formed, but the mechanism seems to be able to eject
 (with relatively low velocity) an amount
 of r-elements sufficient for enrichment of the interstellar matter.

  However, it is not yet clear quantitatively how important
 the possible changes both in the rates $\lambda_{\beta},\; \lambda_{\nu e}$,
 and their ratios could be for the neutrino nucleosynthesis issues.
  These changes are connected with a possible dependence of
  $\lambda_{\beta}$ and $\lambda_{\nu e}$ on temperature and density
  owing to, for example, the contribution of excited nuclear states and
  the Pauli principle for electrons.

\begin{figure}
\vspace{-4.5cm}
\grpicture{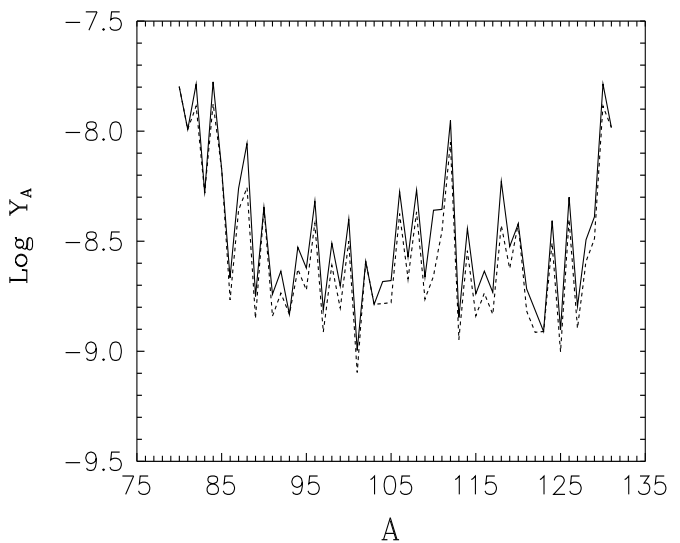}
\vspace{-1cm}
\caption{The influence of the neutrino capture by heavy nuclei
  on the r-process for duration $\tau=10\, $sec
  (the capture is on --- solid line,
   off --- dashed line; $Y_{\protect\indis{n}}(t)$ for Case~6).
}
\label{fig9n}
\end{figure}

 The conditions chosen by us, in particular
 $R=\xmn{5}{8}$cm, appear to be privileged in a certain sense.
 Here the supernova neutrino pulse leads to a neutron yield already
 sufficient to support the \mbox{r-process} through  helium breakup,
 but it is not yet adequate for the {\em direct\/}
 formation of heavy nuclei during the r-process
 due to an increase in the neutron excess by
 the electron neutrino absorption and
 the neutrino-induced evaporation of neutrons from heavy nuclei
 (Domogatsky \& Nadyozhin, 1978).
 Figure~\ref{fig9n} shows the influence of the neutrino capture
 by heavy nuclei on the r-process.
 In order to single out this effect,
 the emission of beta-delayed neutrons was not included in these two runs.
 We see only an acceleration of the r-process but it is hard to find
 any additional smoothing of the odd-even effect.

 When $R$ decreases, the rate $\lambda_{\nu e}$
 increases, and the growth of the element synthesis due to
 neutrino capture relative to the beta-decays at the waiting points
 can lead finally to smoothing of the element yield curve.
 This issue, however, needs further investigation.

 Recently, it was shown that the extremely metal poor stars
 are made of matter enriched with heavy elements synthesized in
 a well developed r-process (Cowan et al., 1996a, 1996b;
 Ryan et~al., 1996; Sneden et al., 1996).
 In the near future, we are going to extend our calculations to
 metallicities as low as Z$=0.001\mbox{Z}_{\textstyle\odot}$ and
 even less to estimate the resulting increase in $R_\indis{cr}$.

 \vspace{0.5cm}

 In general, we conclude that it is very difficult to obtain the
 neutrino-induced r-process in a helium shell at radii
 $R>R_\indis{cr}\approx 10^9$cm,
 in a good agreement with Woosley et~al. (1990).
 The current presupernova models have radii of the helium shell at
 least several times larger than $R_\indis{cr}$ specified above.
 Nevertheless, we continue to search for the  physical conditions
 appropriate for the neutrino-driven r-process, not binding
 ourselves with the question of compatibility
 with the current presupernova models and having in mind probable
 modifications of the models involved, in particular, large scale
 convection, rotation, and magnetic fields.

\begin{acknowledgements}
 Part of this work was done
 during the stay of D.K.N. in the Astronomical Institute University
 of Basel and kindly supported by Swiss National Science Foundation.
 It is a great pleasure for D.K.N. to acknowledge the support and
 to thank Prof. G.A. Tammann for heartfelt hospitality.

   I.V.P. also gratefully acknowledges the hospitality
 of the  MPA and Max Planck Gesellschaft for the opportunity
  to finish this work and fulfill numerical computations
 at  MPI f\"ur Astrophysik.
  We are grateful to Prof. F.-K. Thielemann for providing us with his
 database of thermonuclear reaction rates and
 to S.E. Woosley and A.M. Bykov for stimulating discussions.
 The work was also supported
 by grants Nos. 96-02-17604 and 96-02-16352 of Russian Foundation of
 Fundamental Research  (RFFI) and ISTC Project No 370-97.

 Our special thanks to B.S. Meyer (the Referee) whose constructive critics
 prompted us to improve both the text and contents of the paper.
\end{acknowledgements}

\vspace*{3mm}

\noindent {\bf Appendix: a selfconsistent solution for $Y_\indis{n}$}\\

\vspace*{1mm}

\noindent  We use two nuclear kinetic codes.
 One deals with 101 reactions involving 24 light nuclides from \isn{H}{1}
 and \mbox{ n } through \isn{F}{18} (Code~L hereafter), listed in
 Tables 1 and 2. In addition, it takes
 into account the absorption of neutrons and protons by \isn{Fe}{56}.
 Another code (Code~H), designed for studying the r-process, connects
 more than 3000 heavier nuclides in a network of neutron-capture reactions
 and beta-decays. The two codes are physically connected through
 the exchange of neutrons (the proton-capture by Fe in Code~L is virtually
 unimportant at rather low temperatures considered here). The r-process
 converts initial Fe into heavier nuclei which continue to absorb neutrons
 at a rate that changes as the r-process nucleosynthesis "wave" reaches
 successively heavier and heavier nuclei. Figure \ref{nnoht1}
 shows the results of various assumptions concerning the rate of the neutron
 absorption in the r-process, made in Code~L, to supply Code~H with the
 neutron abundance $Y_\indis{n}(t)$ that would permit to calculate
 the r-process without appealing to Code~L any more. Case~6 was chosen
 as a compromise.

 There are two ways to check the accuracy of such an approximation.
 The most straightforward way would be to unify the two networks in
 one code. This time-consuming work has been planned for the future.
 Another way is to iterate $Y_\indis{n}(t)$ function as described below.

 First, one calculates the r-process using an approximate solution for
 time-dependent neutron density $Y^0_\indis{n}(t)$ from Code~L
 instead of \ewi{Yndt}.
 The calculations provide the rate of the neutron density change, required
 by the r-process $\dot Y^0_\indis{n}(t)$ and given by the right-hand side
 of \ewi{Yndt}. Then, using this
 $\dot Y^0_\indis{n}$ one begins to run Code~L with slightly modified
 differential equation for $Y_\indis{n}$:
 \begin{equation}
  dY_\indis{n}/dt\; =\; B\, +\, Y_\indis{n}\, f^0(t)\, ,
  \label{Ynit}
  \end{equation}
 where  $f^0(t)\equiv\dot Y^0_\indis{n}(t)/Y^0_\indis{n}(t)$ and
 $B$ denotes all the neutron absorbing and emitting reactions given
 in Tables 1 and 2 except for the reaction Fe$(n,\gamma)$.
 In Case~6, the second term in the right-hand side of \ewi{Ynit} equals
 exactly to the unmodified right-hand side of \ewi{dFedt} as described
 in section 4.
 As a result of this  Code~L run one gets an iterated function
 $Y^1_\indis{n}(t)$. Entering Code~H again, now with $Y^1_\indis{n}(t)$,
 one obtains $\dot Y^1_\indis{n}(t)$ that gives the second
 approximation $Y^2_\indis{n}(t)$ for $Y_\indis{n}(t)$
 after running Code~L and so on \ldots

\begin{figure}
\vspace{+20mm}
\epsfysize=20cm
\epsfxsize=17cm
\vspace{-110mm}
\epsffile{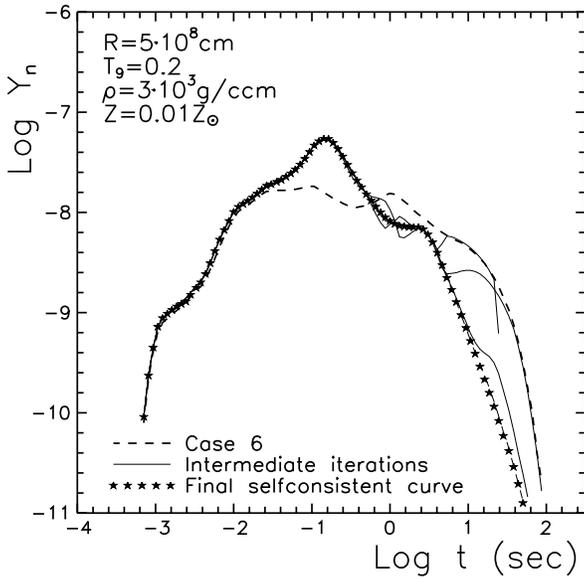}
\vspace{-35mm}
\caption{\noindent The iterative procedure converting the initial
 approximate  $Y_{\protect\indis{n}}(t)$ for Case~6 (dashed line)
 into a final selfconsistent solution (stars).}
\label{selfYn}
\end{figure}

 Figure \ref{selfYn} shows the results of such an iterative fairly
 fast-converging procedure.
 It takes 4--5 iterations to get the final selfconsistent solution
 shown  in \fref{selfYn}  with stars nonconnected by line. The thin line
 underlying the stars represents the last but one iteration that
 is almost indiscernible from the final solution.

\begin{figure}
 \vspace{-105mm}
\grpicture{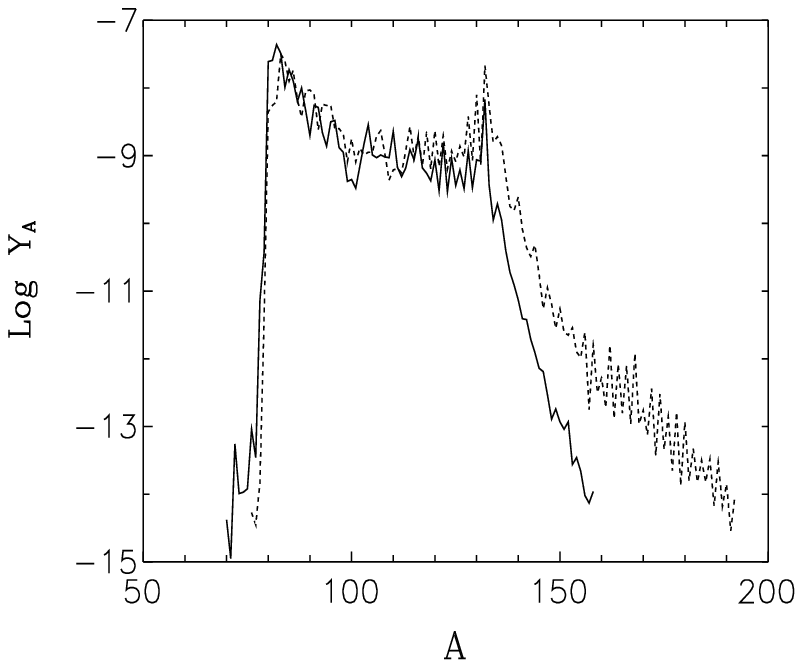}
\vspace{50mm}
\caption{\noindent Same as Version I in \fref{fig3a} but for
 selfconsistent $Y_{\protect\indis{n}}(t)$ from \fref{selfYn}.}
\label{fig13a}
\end{figure}

 From \fref{selfYn}, one can conclude that Case~6 ensures a reasonable
 approximation for $Y_\indis{n}(t)$ as long as $t\lsim 5\,$s. However,
 for longer time ($t\gsim 10\,$s), Case~6 overestimates $Y_\indis{n}(t)$
 by an order of magnitude --- the r-process consumes about 90\% of
 neutrons provided by the neutrino breakup of \isn{He}{4}.
 We remind that in Case~6 initial Fe proves to be totally burnt
 by $t\approx 1\,$s. Therefore for longer times, Case~6 actually neglects
 the absorption of neutrons by the r-process. Using the selfconsistent
 solution for $Y_\indis{n}(t)$ (stars in \fref{selfYn}) instead of that
 for Case~6 (dashed curve in \fref{selfYn}), we have recalculated the
 r-process for times $t=10$ and $35\,$s as shown in \fref{fig13a}.
 Although the r-process yields changed at $t=10\,$s but not so
 significantly as to alter our conclusions in the main text of the paper.
 At $t=35\,$s, the changes are crucial ---
 owing to a shortage in neutrons, the r-process gets virtually exhausted
 and fails to form the abundance peak at $A \approx 200$. To make
 the r-process as far as $A=200$ and beyond under physical
 conditions discussed here ($T_9=0.2, R=\xmn{5}{8}$cm) one needs
 apply to lower metallicities (Z$ \, < \, 0.01\,\mbox{Z}_{\textstyle\odot}$).
 For the selfconsistent solution, the total number of neutrons
 consumed in the r-process per seed \isn{Fe}{56} nucleus is equal to
 about 50.

\end{document}

%% file: h0434_in.tex
%
\begin{table}[htb]
\begin{center}
 Table 1. The list of nuclear reactions

\vspace{0.2cm}

\begin{tabular}{lll}
\hline\hline
         &               &      \\[-0.32cm]
 $^1$H(n,$\gamma$)$^2$H + inv.  & $^7$Li($\alpha$,$\gamma$)$^{11}$B
                                & $^{13}$C(n,$\gamma$)$^{14}$C\\
 $^2$H(p,$\gamma$)$^3$He + inv. & $^7$Li($\alpha$,n)$^{10}$B + inv.
                                & $^{13}$C($\alpha$,n)$^{16}$O\\
 $^2$H(p,n)2$^1$H               & $^7$Be(p,$\gamma$)$^8$B + inv.
                                & $^{14}$C(p,n)$^{14}$N \\
 $^2$H(D,p)$^3$H                & $^7$Be(D,p)2$^4$He
                                & \qquad + inv.\\
 $^2$H(D,n)$^3$He               & $^7$Be(T,np)2$^4$He
                                & $^{14}$C(p,$\gamma$)$^{15}$N\\
 $^3$H(p,$\gamma$)$^4$He        & $^7$Be($^3$He,2p)2$^4$He
                                & $^{13}$N(p,$\gamma$)$^{14}$O\\
 $^3$H(p,n)$^3$He + inv.        & $^7$Be($\alpha$,$\gamma$)$^{11}$C
                                & $^{13}$N(n,$\gamma$)$^{14}$N\\
 $^3$H(D,n)$^4$He               & $^7$Be($\alpha$,p)$^{10}$B + inv.
                                & $^{13}$N($\alpha$,p)$^{16}$O\\ 
 $^3$H(T,2n)$^4$He              & $^9$Be(p,$\gamma$)$^{10}$B
                                & $^{14}$N(p,$\gamma$)$^{15}$O\\ 
 $^3$He(D,p)$^4$He              & $^9$Be(p,np)2$^4$He $^\indis{a}$
                                & $^{14}$N($\alpha$,$\gamma$)$^{18}$F\\
 $^3$He(T,np)$^4$He             & $^9$Be(p,D)2$^4$He
                                & $^{15}$N(p,$\gamma$)$^{16}$O\\
 $^3$He(T,D)$^4$He              & $^9$Be($\alpha$,n)$^{12}$C $^\indis{b}$
                                & $^{15}$N(p,$\alpha$)$^{12}$C\\  
 $^3$He($^3$He,2p)$^4$He        & $^{10}$B($\alpha$,n)$^{13}$N
                                & $^{14}$O(n,p)$^{14}$N\\
 $^4$He($^3$He,$\gamma$)$^7$Be + inv. & $^{10}$B(p,$\gamma$)$^{11}$C 
                                & $^{15}$O(n,p)$^{15}$N\\
 $^4$He(T,$\gamma$)$^7$Li + inv.& $^{11}$B(p,$\gamma$)$^{12}$C
                                & $^{15}$O(n,$\alpha$)$^{12}$C\\
 $^4$He(D,$\gamma$)$^6$Li + inv.& $^{11}$B(p,n)$^{11}$C + inv.
                                & $^{18}$F(n,$\alpha$)$^{15}$N\\
 $^4$He(T,n)$^6$Li + inv.       & $^{11}$B($\alpha$,p)$^{14}$C + inv.
                                & "Fe"(n,$\gamma$) \\
 $^4$He($^3$He,p)$^6$Li + inv.  & $^{11}$B($\alpha$,n)$^{14}$N + inv. 
                                & "Fe"(p,$\gamma$) \\
 $^4$He($2\alpha,\gamma$)$^{12}$C & $^{11}$B(p,$2\alpha$)$^4$He &\\
 $^4$He($\alpha$n,$\gamma$)$^9$Be + inv.& 
                                  $^{11}$C(p,$\gamma$)$^{12}$N + inv. &\\
 $^6$Li($\alpha$,p)$^9$Be + inv.   & $^{11}$C(n,$\gamma$)$^{12}$C        &\\
 $^6$Li($\alpha$,$\gamma$)$^{10}$B & $^{11}$C(n,$2\alpha$)$^4$He         &\\
 $^6$Li(p,$\gamma$)$^7$Be          & $^{11}$C($\alpha$,p)$^{14}$N + inv. &\\
 $^7$Li(p,n)$^7$Be + inv.          & $^{12}$C(p,$\gamma$)$^{13}$N + inv. &\\
 $^7$Li(p,$\alpha$)$^4$He          & $^{12}$C(n,$\gamma$)$^{13}$C        &\\
 $^7$Li(D,n)2$^4$He   & $^{12}$C($\alpha,\gamma$)$^{16}$O $^\indis{c}$   &\\
 $^7$Li(T,2n)2$^4$He               & $^{13}$C(p,$\gamma$)$^{14}$N        &\\
 $^7$Li($^3$He,np)2$^4$He          & $^{13}$C(p,n)$^{13}$N + inv.&\\[-0.4cm]
                          &        & \\ \hline\hline
\end{tabular}
\end{center}
 $^\indis{a}$ The reaction goes through unstable $^9$B$\rightarrow 2^4$He$+$p
 (life time $\sim$\xmn{1}{-16}s).\\
 $^\indis{b}$ The rate is taken from Wrean et al. (1994).\\
$^\indis{c}$ The rate is multiplied by constant 1.7 (Weaver and Woosley, 1993).
\end{table}
\begin{table}[htb]
\begin{center}
 Table 2. The weak interaction processes

\vspace{0.2cm}

\begin{tabular}{ll}
\hline\hline
                                                    & \\[-0.32cm]
 $^{\phantom{1}4}$He($\nu,\nu^\prime$)$^4$He$^\ast$ &
 $\mu$ and $\tau$ neutrinos and antineutrinos,\\
 $^{\phantom{1}4}$He$^\ast\rightarrow ^3$H\phantom{e}$+$p        &
 the cross sections and branching ratios \\
 $^{\phantom{1}4}$He$^\ast\rightarrow ^3$He$+$n                  &
 from Woosley et al. (1990).\\
 $^{\phantom{1}1}$H$(\tilde{\nu}_e,e^+)$n & \\
 $^{\phantom{1}8}$B$\rightarrow 2^{\phantom{1}4}$He$+e^+$  & 
 $ \tau =0.770\;$s\\
 $^{12}$N$\rightarrow\phantom{2}^{12}$C\phantom{e}$+e^+$   & 
 $ \tau =0.011\;$s\\
 $^{15}$O$\rightarrow\phantom{2}^{15}$N\phantom{e}$+e^+$ & 
 $ \tau = 122\;$s\\
       & The lifetimes $\tau$ from Tuli (1990)\\[-0.4cm]
                                                 & \\ \hline\hline
\end{tabular}
\end{center}
\end{table}